# Wide-field, high-resolution Fourier ptychographic microscopy

Guoan Zheng*, Roarke Horstmeyer, and Changhuei Yang

Electrical Engineering, California Institute of Technology, Pasadena, CA 91125, USA

*Correspondence should be addressed to: <gazheng@caltech.edu>

**Abstract:**
In this article, we report an imaging method, termed Fourier ptychographic microscopy (FPM), which iteratively stitches together a number of variably illuminated, low-resolution intensity images in Fourier space to produce a wide-field, high-resolution complex sample image. By adopting a wavefront correction strategy, the FPM method can also correct for aberrations and digitally extend a microscope's depth-of-focus beyond the physical limitations of its optics. As a demonstration, we built a microscope prototype with a resolution of 0.78 µm, a field-of-view of approximately 120 mm$^2$, and a resolution-invariant depth-of-focus of 0.3 mm (characterized at 632 nm). Gigapixel color images of histology slides verify FPM's successful operation. The reported imaging procedure transforms the general challenge of high-throughput, high-resolution microscopy from one that is coupled to the physical limitations of the system's optics to one that is solvable through computation.

The throughput of an imaging platform is fundamentally limited by its optical system's space-bandwidth product (SBP)[1], defined as the number of degrees of freedom it can extract from an optical signal. The SBP of a conventional microscope platform is typically in megapixels, regardless of its employed magnification factor or numerical aperture (NA). As a reference point, a standard 20× microscope objective (MPLN 20×, 0.4 NA, Olympus) has a resolution of 0.8 µm and a 1.1-mm diameter field-of-view (FOV), corresponding to an SBP of approximately 7 megapixels. Increasing the SBP of a microscope is fundamentally confounded by the scale-dependent geometric aberrations of its optical elements[1], thus resulting in a compromise between achievable image resolution and FOV.

A large SBP in microscopy, however, is highly desired in biomedical applications, such as digital pathology, hematology, immunohistochemistry, and neuroanatomy. The strong need in biomedicine and neuroscience to digitally image large numbers of histology slides for analysis has prompted the commercial development of sophisticated mechanical scanning microscope systems and lensless microscopy setups. Artificially increasing an imaging system's SBP by mechanical means is suboptimal, as it requires precise control over actuation, optical alignment, and motion tracking. Furthermore, a mechanical solution simply accepts the intrinsic resolution limit and SBP of a conventional microscope's optics, neglecting the computationally addressable problem of resolution enhancement. Likewise, lensless microscopy methods, such as digital in-line holography[2,3] and contact-imaging microscopy[4-6], offer unique imaging capabilities but also exhibit certain drawbacks. Whereas optically compact, digital in-line holography works well for sparse samples, contact-imaging microscopy requires a sample to be in close proximity to the sensor.

Here, we present a computational imaging method that is capable of providing a scalable SBP for most existing microscopes without involving mechanical scanning or phase measurements. While the method does require the acquisition of a plurality of images, it does so non-mechanically and accomplishes its SBP improvement using the plural data to overcome the physical limitations of the system's optics.

The imaging method we introduce and demonstrate, termed Fourier ptychographic microscopy (FPM), shares its roots with interferometric synthetic aperture microscopy[7-20], ptychography[21-27], phase retrieval[28-31], light-field imaging[32-35], structured illumination[36], and adaptive optics[37]. It works by iteratively stitching together a number of low-resolution images in Fourier space to recover an accurate high-resolution, high-SBP output image. Unlike systems designed to increase the SBP of a single acquisition[38], combining time-sequential measurements for the same goal allows our setup to maintain a simple and compact form factor. The main design strategy of FPM is similar to that of interferometric synthetic aperture microscopy[7-20]: expanding the SBP in Fourier space through multi-image fusion. However, because no measured phase information is needed for FPM, our setup eliminates the design challenges associated with interferometric detection schemes. Furthermore, the image recovery procedure of FPM follows a strategy common with ptychography (i.e., scanning diffraction microscopy)[21-27]: iteratively solving for a sample estimate that is consistent with many intensity measurements. Unlike ptychography, however, FPM's object support constraints are imposed in the Fourier domain, offering several unique advantages and opportunities.

By adding a simple light emitting diode (LED) matrix illumination module and applying FPM's reconstruction algorithm, we transform a conventional optical microscope into a high-resolution (0.78 µm, 0.5 NA), wide-FOV (~120 mm$^2$) microscope with a final SBP of approximately 1 gigapixel. Our joint optical-digital solution further allows us to exploit adaptive optics-based wavefront correction strategies to compensate for aberrations and expand depth-of-focus beyond conventional optical limits. Specifically, we use our FPM procedure to extend a conventional microscope's 80-µm depth-of-focus to approximately 0.3 mm, creating a platform with a large tolerance to microscope slide placement errors. We will briefly outline FPM's operation and experimental setup, discuss how to apply FPM's digital wavefront correction technique, and demonstrate successful gigapixel imaging of a pathology slide.

**Results**
**Principle of FPM**
The data collection procedure of FPM is straightforward. We place a 2D sample at the focal plane of a low-NA microscope objective and collect a sequence of *N* images, with the sample successively illuminated by plane waves at *N* different angles. As a key distinction from other synthetic aperture techniques, we only acquire intensity images of the sample—no interferometric measurements are needed. The use of a low-NA objective lens allows a large FOV to be captured at the expense of a low spatial resolution. In this section, we assume the sample is located at the in-focus position of the microscope objective. Later, we will demonstrate that computational refocusing of a mispositioned sample is also possible. Based on N collected low-resolution intensity images, we computationally reconstruct a high-resolution image of the sample following the recovery procedure shown in Figure 1.

Before explaining the procedure, we first note that our recovery process alternates between the spatial $(x - y)$ and Fourier $(k_x - k_y)$ domains, where $k$ represents wavenumber. Second, we assume that illuminating a thin sample by an oblique plane wave with a wave vector $(k_x, k_y)$ is equivalent to shifting the center of the sample's spectrum by $(k_x, k_y)$ in the Fourier domain. Third, we assume our objective lens's filtering function (i.e., coherent optical transfer function) in Fourier space is a circular pupil with a radius of NA*$k_0$, where $k_0 = 2π/λ$ is the wavenumber in a vacuum.

FPM generates a high-resolution image $I_h$ from a set of N low-resolution measurements, $I_{lm}(k_x^i, k_y^i)$ (indexed by their illumination wave vector $k_x^i$, $k_y^i$, with $i$ = 1, 2…N) as follows. Subscripts '$h$', '$l$', and '$m$' denote high-resolution, low-resolution, and measurement, respectively. First, the FPM method starts by making an initial guess of the high-resolution object function in the spatial domain, $\sqrt{I_h}e^{i\varphi_h}$. A good starting point is to select $\varphi_h$=0 and $I_h$ as any up-sampled low-resolution image (an initial guess with constant value also works). The Fourier transform of the initial guess creates a broad spectrum in the Fourier domain (Fig. 1, left).

Second, we select a small subregion of this spectrum, equivalent to a low-pass filter, and apply Fourier transformation to generate a new low-resolution target image $\sqrt{I_l}e^{i\varphi_l}$. The applied low-pass filter shape is a circular pupil, given by the coherent transfer function of the objective lens. The position of the low-pass filter is selected to correspond to a particular angle of illumination. For example, the subregion enclosed by the red circle in Figure 1 corresponds to an image collected under normally incident illumination ($k_x^1 = 0$, $k_y^1 = 0$).

Third, following phase retrieval concepts developed by Fienup[28-31], we replace the target image's amplitude component $\sqrt{I_l}$ with the square root of the low-resolution measurement obtained under illumination angle $i$, $\sqrt{I_{lm}}$, to form an updated, low-resolution target image $\sqrt{I_{lm}}e^{i\varphi_l}$. We then apply Fourier transformation to this updated target $\sqrt{I_{lm}}e^{i\varphi_l}$ and replace its corresponding subregion of the high-resolution Fourier space. In other words, for $i$ = 1, we update the area enclosed by the red circle in Figure 1 with image $I_{lm}(k_x^1, k_y^1)$, where $k_x^1 = 0$, $k_y^1 = 0$.

Fourth, for the $i^{th}$-shifted subregion, we repeat steps 2 and 3 (select a small, circular region of k-space and update it with measured image data). Other examples are represented by the green and blue circles in Figure 1. Each shifted subregion corresponds to a unique, low-resolution intensity measurement $I_{lm}(k_x^i, k_y^i)$, and each subregion must overlap with neighboring subregions to assure convergence. This data redundancy requirement is also present in ptychography[22,39]. This iterative update continues for all N images, at which point the entire high-resolution image in Fourier space has been modified with data from all low-resolution intensity measurements.

Lastly, steps 2–4 are repeated until a self-consistent solution is achieved (for the simulation and experimental data shown in this paper, we only repeat these steps once). At the end of this iterative recovery process, the converged solution in Fourier space is transformed to the spatial domain to recover a high-resolution field $\sqrt{I_h}e^{i\varphi_h}$, offering an accurate image of the targeted 2D sample (Fig. 1, right) with a dramatically increased SBP (high-resolution with wide-FOV). A discussion of the computational cost of the above recovery procedure can be found in Supplementary Note 1. We also performed a set of numerical simulations to validate the proposed FPM method in Supplementary Note 2.

Drawing connections and distinctions between this iterative process and two related modalities, light-field imaging[32-35] and ptychography[21-27], helps clarify FPM's principle of operation. A light-field microscope uses a microlens array at its image plane to project $M$ subimages onto its sensor. By extracting different pixels from each subimage, different perspective views can be synthesized, each corresponding to a small aperture of the objective's pupil plane[32,33]. Similar to a light-field microscope, the FPM setup also captures multiple perspective images of a sample, corresponding to different, small apertures in the Fourier domain. However, three key differences allow the FPM to create a high-resolution output. First, FPM records larger scattering angles than does a standard light-field microscope. Second, light-field microscopes sacrifice spatial resolution to acquire all perspective images in one single snapshot[32,33], whereas FPM acquires each perspective over time. Third, light-field microscopes use the interplay between spatial and angular resolvability to achieve refocusing through a 3D sample, whereas FPM applies this interplay to achieve a different goal: different angular perspectives are synthesized to increase a 2D object's spatial resolution. We also note that an FPM dataset of a 3D object can be processed in a similar way as a light-field microscope to achieve 3D sample refocusing and rendering[40].

Ptychography[21-27] is a lensless imaging method originally proposed for transmission electron microscopy and brought to fruition by Faulkner and Rodenburg with the introduction of transverse translation diversity[22,39]. The basic idea of ptychography is to illuminate a sample with a focused beam and repeatedly record its far-field diffraction pattern as a function of sample position. Iterative retrieval methods are then applied to invert the diffraction process and recover the sample's amplitude and phase from this set of measurements. It is clear that FPM and ptychography both iteratively seek a complex field solution that is consistent with many intensity measurements. With ptychography, the object support for phase retrieval is provided by the confined illumination probe in the spatial domain; therefore, the sample (or the probe) must be mechanically scanned through the desired FOV. With FPM, however, the object support is provided by the confined NA in the Fourier domain (a circular pupil). In this regard, FPM appears as the Fourier counterpart of ptychography, justifying the proposed name. By imposing object support in the Fourier domain, FPM naturally offers a large, fixed FOV, a higher signal-to-noise ratio (with focusing elements), and no mechanical scanning as compared to conventional ptychography. Furthermore, as discussed in below, FPM can also digitally correct for aberrations common to simple low-NA focusing elements.

**Experimental setup and characterization**
To experimentally validate the FPM method, we used an Olympus BX 41 microscope, a 2× apochromatic objective lens (Plan APO, 0.08 NA, Olympus), and an interline CCD camera (Kodak KAI-29050, 5.5-μm pixel size) as our experimental setup. We then introduced a programmable color LED matrix placed approximately 8 cm under the sample stage as a variable illumination source (Fig. 2a and 2b, also refer to Methods).

Resolution improvement provided by the FPM method is demonstrated with a USAF resolution target imaging experiment in Figure 2c and 2d (also refer to Supplementary Video 1). Figure 2c1 shows a full-FOV raw intensity image acquired by the FPM platform. Figure 2c2 shows a magnified view of the raw data, with a pixel size of 2.75 μm at the object plane (CCD pixel size divided by the magnification factor). The corresponding high-resolution FPM reconstruction is shown in Figure 2d for comparison, with a maximum synthetic NA is 0.5 set by the maximum angle between the optical axis and an LED. In our FPM reconstruction, the feature of group 9, element 3 on the USAF target (0.78 μm line width) is clearly resolved. This verifies our prototype platform's expected synthetic NA of 0.5, following the Rayleigh criterion (refer to Fig. S2 for FPM reconstructions with different synthetic NAs and Fig. S5 for reconstructed image line traces). In Figure S3, we further determine the depth-of-focus of the proposed platform to be approximately 100 μm without any computational correction applied, which is approximately equal to the 80 μm depth-of-focus associated with the 2× objective used in the experiment, but is approximately 25 fold longer than that of a conventional microscope objective with a similar 0.5 NA.

**Digital wavefront correction**

While the FPM method does not require phase information as input, its operation implicitly accommodates phase during iterative reconstruction. As we will demonstrate, the depth-of-focus of our FPM prototype can be significantly extended beyond that of the employed objective lens using a numerical strategy to compensate for aberrations in the pupil function[19,41].

This digital correction process is inspired by similar wavefront correction concepts in adaptive optics[37]. The basic idea is to digitally introduce a phase map to our coherent optical transfer function to compensate for aberrations at the pupil plane during the iterative image recovery process. The FPM algorithm incorporates this compensation into two additional multiplication steps (steps 2 and 5 in Fig. 3a and 3b). Specifically, step 2 models the connection between the actual sample profile and the captured intensity data (with included aberrations) through multiplication with a pupil function $e^{i\cdot\varphi(k_x,k_y)}$, whereas step 5 inverts such a connection to achieve an aberration-free reconstructed image. Sample defocus is essentially equivalent to introducing a second-order Zernike mode, or a quadratic phase factor, to the pupil plane (i.e., a defocus aberration[42]):

$$e^{i\cdot\varphi(k_x,k_y)} = e^{i\sqrt{(2\pi/\lambda)^2-k_x^2-k_y^2}\cdot z_0}, \; k_x^2+k_y^2 < (NA\cdot 2\pi/\lambda)^2, \quad (1)$$

where $k_x$ and $k_y$ are the wave numbers at the pupil plane, $z_0$ is the defocus distance, and $NA$ is the objective's numerical aperture.

Simulations of the proposed digital wavefront correction strategy are provided in Supplementary Note 2, whereas Figure 3 experimentally demonstrates FPM's ability to fully resolve an object given a set of intensity images defocused by 150 μm. The significance of wavefront correction is made clear by comparing reconstruction results without (Fig. 3d) or with (Fig. 3e) digital addition of a defocused pupil. We note that, in Figure 3e, the defocus distance is known a priori. If the defocus distance is unknown, we can digitally adjust the 'z' parameter to different values, reconstruct the corresponding FPM images, and pick the sharpest image through observation or by a computer algorithm. This approach can be extended for tilted samples as well. Here, we digitally adjust the 'z' parameter to achieve acuity for each vignette of the whole image and combine the in-focus vignette to form a focused image of the tilted sample. From Figure 3, we conclude that our FPM prototype can achieve a resolution-invariant depth-of-focus of approximately 0.3 mm with digital wavefront correction (Fig. S5). In contrast, the natural depth-of-focus of the employed 2× objective lens (0.08 NA) is approximately 80 µm. The improvement is even more remarkable if compared to an objective lens with a resolution-matching 0.5 NA, where the FPM prototype's 0.3 mm depth-of-focus offers an approximate factor of improvement of 75.

Finally, we note that alternate digital multiplicative phase factors can be included in steps 2 and 5 to correct for a variety of aberrations, as long as they correctly model the employed optics. Following this strategy, we also correct for astigmatism aberrations of our prototype's objective lens (refer to Methods and Supplementary Note 4 for details on pupil function measurement). In Figure S6, we establish that the effective FOV of our prototype is approximately 1.25 cm in diameter (~120 mm$^2$).

**Gigapixel color imaging of histology slides**
Color FPM images can be created by simply combining results from red, green, and blue LED illumination into each corresponding color channel. We demonstrate color FPM with our prototype by acquiring a wide-FOV color image of a pathology slide (human adenocarcinoma of breast section, Carolina), as shown in Figure 4. Vignette high-resolution views are provided in Figure 4b–4d with a reconstructed pixel size of 0.275 µm. The imaging FOV is approximately 120 mm$^2$, the same as that from a 2× objective (Plan APO, 0.08 NA, Olympus), whereas the maximum achieved NA is 0.5, similar to that of a typical 20× objective (MPLN, 0.4 NA, Olympus). The conventional microscope images taken with 20× and 2× lenses are shown for comparison in Figure 4c2 and 4c3. In Figure S7, we include a detailed comparison between raw data, FPM reconstruction, and a conventional microscope image for two samples: a pathology slide and a blood smear.

The demonstrated SBP of our FPM prototype is approximately 0.9 gigapixels (120 mm$^2$ FOV divided by 0.37$^2$ µm$^2$ Nyquist pixel area, characterized at 632 nm wavelength; refer to Supplementary Note 3). Such a SBP is orders of magnitude larger than that of its constituent 2× objective (16 megapixels) and that of a typical 20× objective (7 megapixels). From another perspective, our FPM prototype can be considered a microscope that combines the FOV advantage of a 2× objective with the resolution advantage of a 20× objective.

**Discussion**
We have demonstrated a simple and cost-effective microscopy imaging method, termed Fourier ptychographic microscopy (FPM). This computation-based method is capable of providing a scalable SBP for most conventional microscopes without requiring mechanical scanning. Fundamentally, it transforms the general challenge of high-throughput microscopy from one that is coupled to the physical limitations of optics to one that is solvable through computation. FPM can be applied to most standard digital microscopes through retrofitting with a readily available

LED matrix. Our discussion has focused on generating a high-SBP intensity image; the capability of FPM-enabled phase imaging will be detailed in future work.

Our FPM prototype has not been optimized for performance speed. At present, our imaging speed is limited by the low illumination intensities provided by the LEDs at the array's edges to the sample. This issue can be addressed by either angling the LEDs inwards or using higher power LEDs. The processing speed can be significantly improved by employing a GPU, as the algorithm process is highly parallelizable. The FPM method requires an overlap of the spectrum in Fourier domains encompassed by each raw image (~60% spectrum overlap in our implementation). This redundancy is necessary as it promotes fast image convergence. It would be worth exploring the exact relationship between data redundancy and convergence in the future. In our current FPM method, we assume that samples are effectively two-dimensional. We believe there are FPM variants that can be developed to handle extended samples. We would also like to reiterate that the current FPM method is not a fluorescence technique, as fluorescent emission profiles would remain unchanged under angle-varied illuminations. However, we believe that it is possible to use patterned illumination with FPM variants to increase the SBP of a fluorescence image. Finally, we believe that higher order aberration corrections and more accurate back aperture characterizations would be worth implementing in future systems to improve the image quality of the FPM reconstruction.

FPM's ability to significantly increase the SBP of a conventional microscope is highly useful for addressing the wide-FOV imaging needs that dominate digital pathology and neuroscience. Furthermore, FPM's digital wavefront correction procedure lends extra flexibility to many biomedicine experiments, by largely eliminating the need to maintain a precise working distance between the sample slide and the microscope objective.

However, we believe that FPM is potentially even more broadly transformative. Conventionally, the quality of an imaging system is largely defined by the physical limitations of its optical elements. For example, spatial resolution is generally characterized by the sharpness of the optical system's point-spread function. The proposed FPM method reduces the optical system to a filtering transfer function of the complex field employed in an iterative recovery process, through which the characteristics of this complex optical transfer function are rendered nominally irrelevant. As long as the low-pass pupil function is accurately characterized, this link between the actual sample profile and captured data may iteratively improve image resolution. It is this underlying robustness that allows our FPM prototype to render high-resolution images with a low-NA objective which is conventionally incapable of optically providing such a narrow point-spread function and long depth-of-focus.

More broadly speaking, FPM can be potentially applied to systems with severe but known aberrations to render high-quality images. Our demonstration of digital wavefront correction provides a viable strategy in this respect. We believe that the development of a general aberration correction procedure using our iterative complex field recovery strategy would be very interesting and useful. Additionally, it can potentially significantly improve X-ray and THz imaging setups that are generally limited by poor and aberrative focusing elements.

**Methods**
**Experimental setup**

The measured distance between the sample stage and the LED array was about 8 cm, and the measured working distance of the objective lens was about 6 mm. The LED matrix contains 32*32 surface mounted, full-color LEDs (SMD 3528), and the lateral distance between two adjacent LEDs is 4 mm. The central wavelengths of the full-color LED are 632 nm (red), 532 nm (green), and 472 nm (blue), each offering an approximately spatially coherent quasi-monochromatic source with an approximate 15 nm bandwidth.

We used an Atmel ATMEGA-328 microcontroller to provide the logical control for the LED matrix. To achieve maximum brightness, the matrix was driven statically rather than in normal scanning mode, eliminating the duty cycle and boosting currents through the LEDs at a maximum level. The measured light intensities were 0.7, 1.0, and 0.4 W/m$^2$ for the red, green, and blue colors, respectively. Measured intensities of different individual LEDs were also used to normalize each corresponding intensity image.

**Image acquisition and reconstruction**

In all figures shown, variable pixel gain was removed by flat-field correction, and hot pixels were identified and removed by interpolation. The sampling requirement of raw images is λ/(2·$NA_{obj}$), where $NA_{obj}$ denotes the NA of the employed objective lens (refer to Supplementary Note 3). To reconstruct a high-resolution image with a maximum synthetic NA of 0.5, we use 137 LEDs for illumination (each LED corresponds to a circle in Fig. S2c2). Due to low light intensities of the LEDs, a long exposure time is required by our prototype, limiting the speed of image acquisition. For the central 49 (7 by 7) LEDs, we acquired three images with three different exposure times (0.005 s, 0.1 s, and 0.5 s), and combined them to obtain a 14-bit high-dynamic range (HDR) image for FPM reconstruction. For LEDs outside this central area, we acquired two images with two different exposure times (0.1 s and 0.8 s) to create an 11-bit HDR image. The HDR combination process is used to suppress the saturation error caused by the overexposed pixels[43]. The total acquisition time for the current prototype is about 3 minutes. With a brighter LED matrix, the maximum throughput will ultimately be determined by the sensor's data transfer rate. For example, using a commercially available 53 fps full-frame camera (VC-25MX, Vieworks), an acquisition time of several seconds can be achieved for a gigapixel image.

During the reconstruction process, we divided each full FOV raw image (5280 × 4380 pixels) into smaller image segments (150 × 150 pixels each). Each set of image segments was then independently processed by the FPM recovery procedure to create a high-resolution image segment (1500 × 1500 pixels). Finally, all high-resolution image segments were combined into one full FOV, high-resolution image (Fig. S8). The benefits of dividing the raw image into smaller segments include the following:

1) Each segment of the raw image can be processed independently, a requirement for parallel computing.

2) Memory requirements for computation are reduced.

3) The light from each LED can be accurately treated as a plane wave for each image segment of the raw image. The incident wave vector for each segment can be expressed as

$$(k_x^i, k_y^i) = \frac{2\pi}{\lambda}\left(\frac{(x_c-x_i)}{\sqrt{(x_c-x_i)^2+(x_c-x_i)^2+h^2}}, \frac{(y_c-y_i)}{\sqrt{(y_c-y_i)^2+(y_c-y_i)^2+h^2}}\right),$$

where $(x_c, y_c)$ is the central position of each small segment of the raw image, $(x_i, y_i)$ is the position of the $i^{th}$ LED, and h is the distance between the LED matrix and the sample.

4) Each small portion can be assigned a specific aberration-correcting pupil function, a common strategy used in wide field imaging[44] (see Supplementary Note 4 for pupil function measurement details).

Using a personal computer with an Intel i7 CPU (no GPU), the processing time for each high-resolution image segment (converting 150 × 150 raw pixels to 1500 × 1500 pixels) is about 2 seconds in Matlab. The total processing time for creating a final full FOV image is about 10 minutes. For color imaging via FPM, we acquired the red, green, and blue channels using their corresponding color LEDs, processing each channel independently. Thus, the total acquisition and processing time for a color image must be multiplied by a factor of 3.

**Acknowledgements**
We are grateful for the constructive discussions and generous help from Mr. Xiaoze Ou, Ms. Ying Min Wang and Mr. Chris Kolner from Caltech. We acknowledge funding from NIH 1DP2OD007307-01.

**Author contributions:** G.Z. conceived the initial idea, designed and implemented the experiment. G.Z., R.H., and C.Y. developed and refined the concept, and wrote the paper.

**Figure 1:**

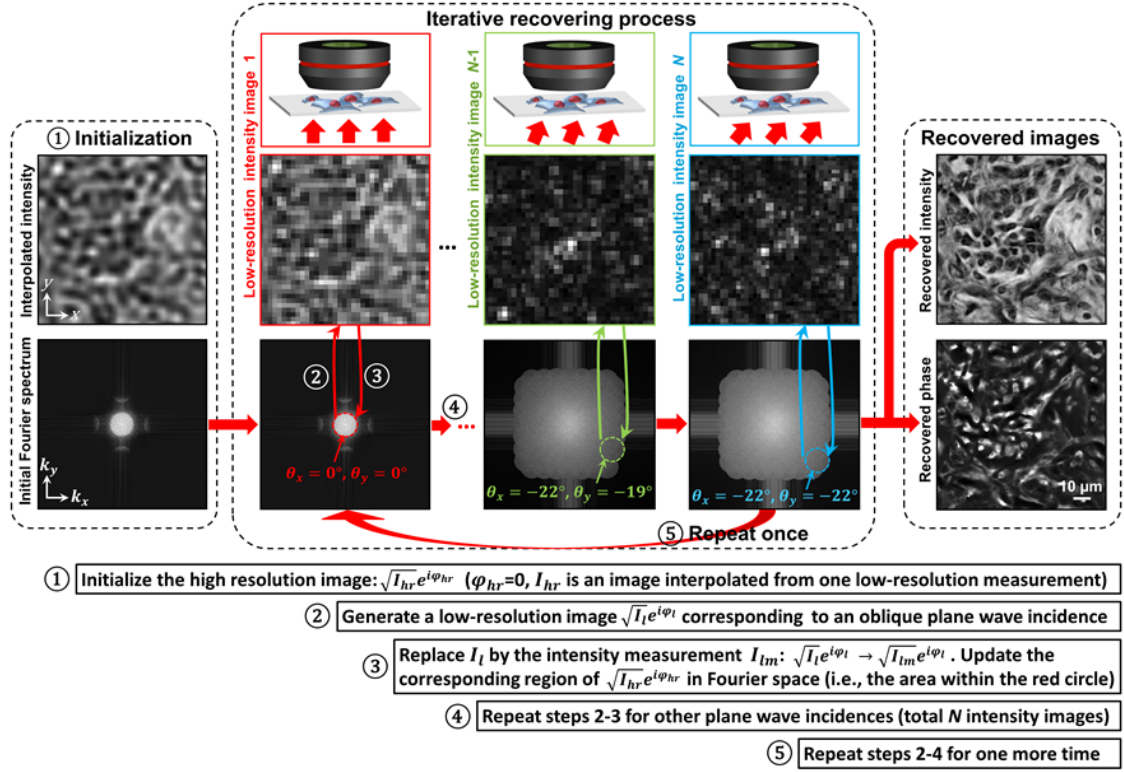

**Fig. 1 FPM's iterative recovery procedure.** Steps 1–5 illustrate FPM's algorithm, following principles from phase retrieval. *N* low-resolution intensity images captured under variable illumination are used to recover one high-resolution intensity image and one high-resolution phase map.

**Figure 2:**

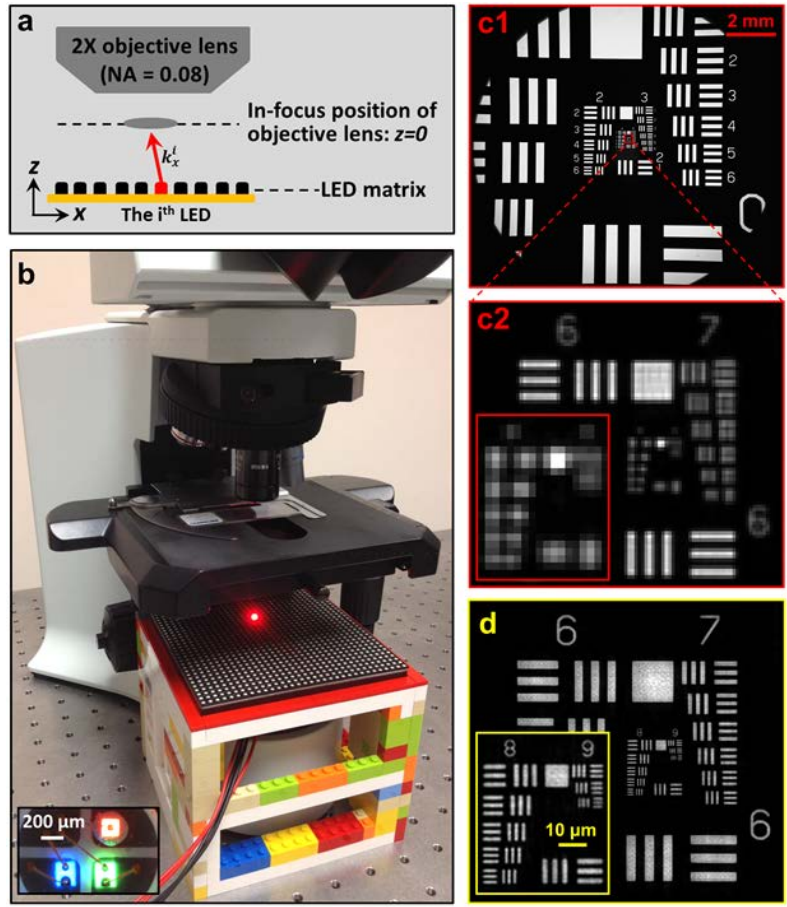

**Fig. 2 FPM prototype setup.** (a) Diagram of the setup. A programmable LED matrix is placed beneath the sample. The i[th] LED illuminates the sample with wave-vector $k_x^i$. (b) The LED matrix and microscope used in experiment, where (Inset) each LED can provide red, green, and blue narrow-band illumination. (c1) A full-FOV raw image of a USAF resolution target. (c2) A magnified view of the raw image, exhibiting a pixel size of 2.75 µm. (d) Our FPM reconstruction of the same region, where we achieve a reconstructed pixel size of 0.275 µm (refer to the discussion of FPM's sampling requirement in Supplementary Note 3). In this reconstruction, the corresponding maximum synthetic NA of the reconstructed image is 0.5, set by the maximum angle between the optical axis and an LED. The entire recovery process is demonstrated in Supplementary Video 1.

**Figure 3:**

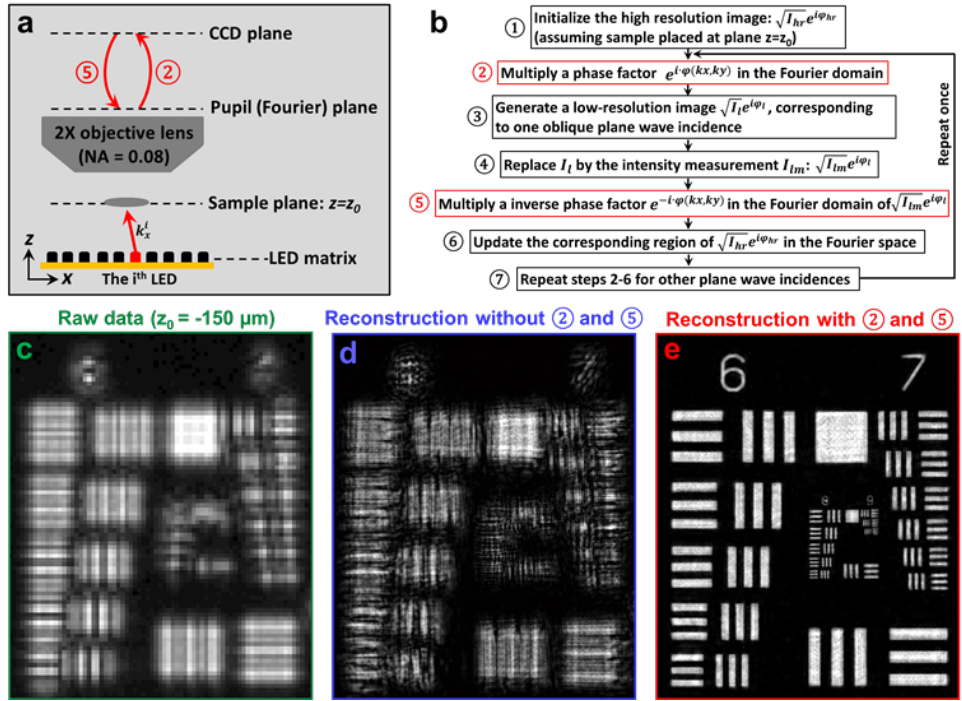

**Fig. 3 Extending depth-of-focus with digital wavefront correction.** (a) The principle of FPM's digital wavefront correction technique. A digital pupil function is introduced in steps 2 and 5 to model the connection between the actual sample profile and the captured intensity data, which may exhibit aberrations caused by defocus. (b) Diagram of FPM's iterative recovery algorithm with the addition of digital wavefront correction. (c) One raw low-resolution image of the USAF target placed at $z_0 = -150$ µm. High-resolution FPM reconstructions without (d) and with (e) steps 2 and 5 added to the iterative recovery procedure.

**Figure 4:**

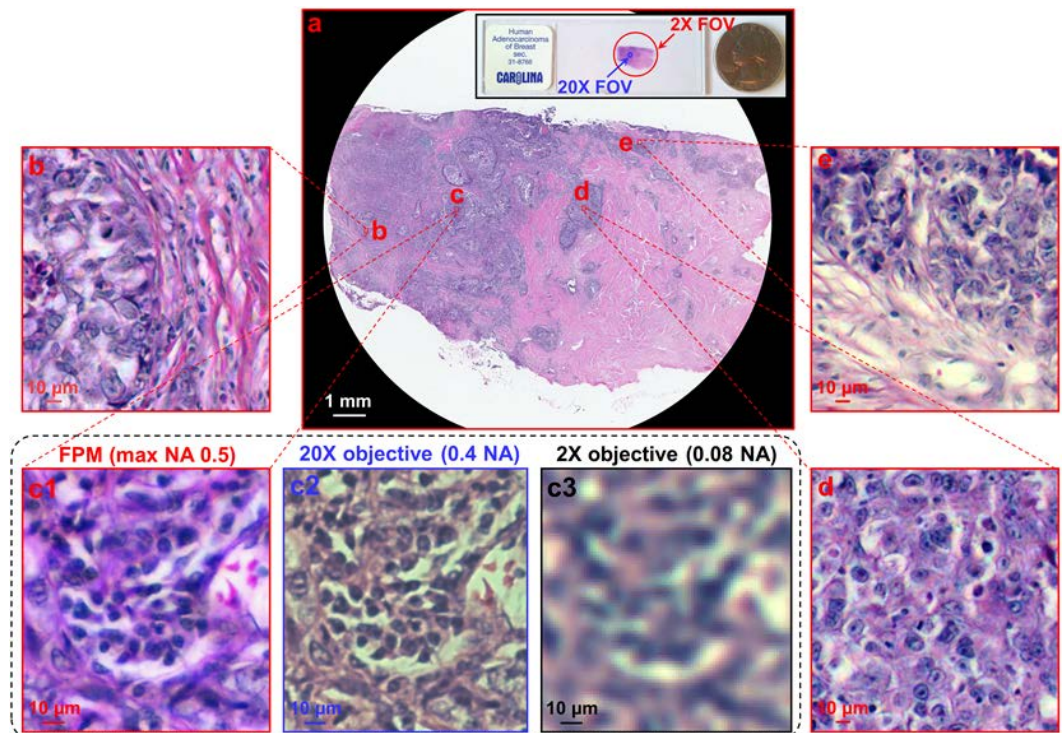

**Fig. 4 Gigapixel color imaging via FPM.** (a) A wide-FOV color image of a pathology slide, with a SBP of approximately 0.9 gigapixels. (b, c1, d, and e): Vignette high-resolution views of the image in (a). Images taken by a conventional microscope with a 20× (c2) and a 2× (c3) objective lens, for comparison. A color image sensor (DFK 61BUC02, Image Source Inc.) is used for capturing (c2 and c3).